 \documentclass[11pt]{article}

\usepackage{amsfonts,amssymb,amsthm}

\usepackage{amsmath}
\usepackage{multicol}
\usepackage{amscd}
\usepackage{hyperref} 
\usepackage{graphics}
\usepackage{graphicx}
\usepackage[applemac]{inputenc}

\usepackage{tikz}
\usepackage{multirow}

\usepackage{hyperref}
\hypersetup{
   colorlinks   =  true,
    linkcolor    = cyan,
    citecolor    = red,
     urlcolor	=magenta,     
}

\usepackage{cite}

\def\be{\begin{equation}}
\def\ee{\end{equation}}
\def\bea{\begin{eqnarray}}
\def\eea{\end{eqnarray}}

\def\>#1{{\mathbf #1}}                 

\def\producto{\cdot}                 

\def\adsw{(A)dS\ }

\def\L{\Lambda}

\def\r{{\eta}}

\def\1{\'{\i}}
\def\R{{\cal R}}
\def\>#1{{\bf #1}}

\begin{document}
%
%
%

%
%

\ 
\vspace{0.5cm}

\begin{center}

\LARGE{{
Interplay between spacetime curvature, speed of light and quantum deformations of relativistic symmetries
}} \\
\end{center}

\begin{center}

{\sc Angel Ballesteros$^1$, Giulia Gubitosi$^{2,3}$,  Flavio Mercati$^{1}$}

\medskip

{$^1$Departamento de F\'isica, Universidad de Burgos, 
09001 Burgos, Spain}

{$^{2}$ Dipartimento di Fisica ``Ettore Pancini'',
Universit\`a di Napoli Federico II, Napoli, Italy}

{$^{3}$ INFN, Sezione di Napoli}

 \medskip
 
e-mail: {\href{mailto:angelb@ubu.es}{angelb@ubu.es}, \href{mailto:giulia.gubitosi@unina.it}{giulia.gubitosi@unina.it}, \href{mailto:fmercati@ubu.es}{fmercati@ubu.es}}


\end{center}

\medskip

\begin{abstract}
Recent work showed that $\kappa$-deformations can describe the quantum deformation of several relativistic models that have been proposed in the context of quantum gravity phenomenology. Starting from the Poincar\'e algebra of special-relativistic symmetries, one can toggle the curvature parameter $\Lambda$, the Planck scale quantum deformation parameter $\kappa$ and the speed of light parameter $c$ to move to the well-studied $\kappa$-Poincar\'e algebra, the (quantum) (A)dS algebra, the (quantum) Galilei and Carroll algebras and their curved versions.  In this review, we survey the properties and relations of these algebras of relativistic symmetries and their associated noncommutative spacetimes, emphasizing the nontrivial effects of interplay between curvature, quantum deformation and speed of light parameters. 
\end{abstract}

  \medskip 
  
\noindent {\sc Keywords}: quantum groups, Poincar\'e group, (Anti)-de Sitter, Galilei group, Carroll symmetries, curvature, deformation, Planck scale, noncommutative spacetimes, quantum gravity, phenomenology.

\tableofcontents

  \medskip

\section{Introduction}

Deformations of relativistic symmetries have been playing a prominent role in the study of  phenomenologically relevant effects of quantum gravity in a ``non-quantum" and ``non-gravitational" regime, such that both the Planck constant $\hbar$ and the Newton constant $G$ are negligible, but their ratio is not, thus leaving the Planck energy $E_P=\sqrt{\frac{c^5 \hbar}{G}}$  finite \cite{ACFKS1,ACFKS2}. 

In this context, a much studied formalism that provides a rigorous mathematical framework for the deformed symmetry models is that of $\kappa$-deformations \cite{LRNT1991,GKMMK1992,LNR1992fieldtheory,Maslanka1993,Lukierski:1993wxa,Zakrzewski1994poincare,MR1994,Lukierski:2016vah}, which turn the Lie algebra describing the Poincar\'e symmetries of special relativity into a Hopf algebra and where the quantum deformation parameter $\kappa$ is assumed to be of the order of the Planck energy~\cite{Majida}. Despite these models being originally derived as a contraction of the quantum (Anti-)de Sitter algebra in the limit of vanishing cosmological constant $\Lambda$, the great majority of the subsequent work focussed exclusively on the $\Lambda=0$ case. 

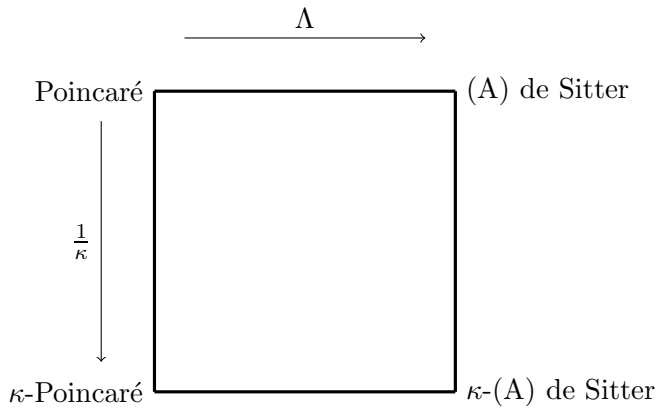
\begin{figure}[htbp]
\begin{center}
\begin{tikzpicture}[thick,scale=4]
 
    \coordinate (B1) at (0.3, -0.3);
    \coordinate (B2) at (1.3, -0.3);
    \coordinate (B3) at (1.3, 0.7);
    \coordinate (B4) at (0.3, 0.7);

    \draw[very thick] (B1) -- (B2);
    \draw[very thick] (B2) -- (B3);
    \draw[very thick] (B1) -- (B4);
    \draw[very thick] (B3) -- (B4);

    \node[anchor=east] at (B1) {$\kappa$-Poincar\'e};
    \node[anchor=west] at (B2) {$\kappa$-(A) de Sitter};
    \node[anchor= west] at (B3) {(A) de Sitter};
    \node[anchor=east] at (B4) {Poincar\'e};

\draw[->,yshift=.2em,thin] (0.4, 0.8)--node[above]{$\Lambda$}(1.2, 0.8);
\draw[->,xshift=-.2em,thin] (0.2,0.6)--node[left]{$\frac 1 \kappa$}(0.2,-0.2);

\end{tikzpicture}
\caption{The various algebras of relativistic symmetries emerging in the regimes set by  different combinations of the cosmological constant $\Lambda$ and the quantum deformation parameter $\kappa$. The arrows point in the direction where the indicated parameter becomes nonzero. We see that the (Anti)-de Sitter algebra and the $\kappa$-Poincar\'e algebra are both deformations of the Poincar\'e algebra, one being a classical deformation and the other a quantum deformation, respectively.}
\label{fig1}
\end{center}
\end{figure}

Nevertheless, some preliminary analyses~\cite{ASS2004,Marciano:2010gq,Bianchi:2011uq,Amelino-Camelia:2012vzf, Rosati:2015pga, Barcaroli:2015eqe} pointed out that nontrivial effects are to be expected due to the interplay between the cosmological constant $\Lambda$ and the quantum deformation parameter $\kappa$, and these effects might have significant implications for  phenomenological analyses that focus on an astrophysical setup where the cosmological expansion is non-negligible \cite{Amelino-Camelia:2020bvx}. This interplay emerges because  the two parameters govern two kinds of deformation of the Poincar\'e algebra, respectively a classical deformation, turning the Poincar\'e algebra into a new Lie algebra describing (Anti-)de Sitter symmetries~\cite{Herranz:2006un}, and a quantum deformation, turning the Poincar\'e algebra into a deformed Hopf algebra (see figure \ref{fig1}). When both deformations are present, the Poincar\'e algebra turns into a $\kappa$-deformed (Anti-)de Sitter Hopf algebra, and novel features emerge, that are governed by products of the two deformation parameters, so that they disappear in both the flat $\Lambda\rightarrow 0$ and the classical $\kappa^{-1}\rightarrow 0$ limits \cite{BHOS1994global,JackkdS,Brunoc,BHM2014plb,BHMN2017kappa3+1,BGH2019spacetime}.

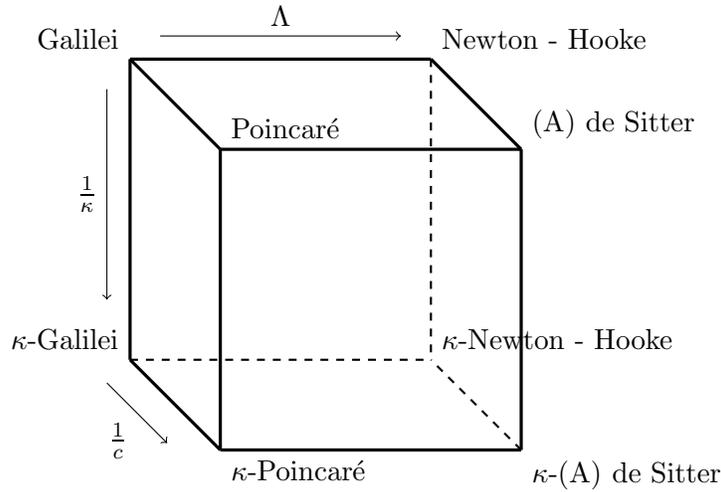
\begin{figure}[htbp]
\begin{center}
\begin{tikzpicture}[thick,scale=4]
    \coordinate (A1) at (0, 0);
    \coordinate (A2) at (0, 1);
    \coordinate (A3) at (1, 1);
    \coordinate (A4) at (1, 0);
    \coordinate (B1) at (0.3, -0.3);
    \coordinate (B2) at (1.3, -0.3);
    \coordinate (B3) at (1.3, 0.7);
    \coordinate (B4) at (0.3, 0.7);

    \draw[very thick] (A1) -- (A2) ;
    \draw[very thick] (A2) -- (A3);
    \draw[dashed] (A3) -- (A4);
    \draw[dashed] (A1) -- (A4);

    \draw[very thick] (A1) -- (B1);
    \draw[very thick] (B1) -- (B2);
    \draw[very thick] (B2) -- (B3);
    \draw[very thick] (B3) -- (A3);
    \draw[dashed] (B2) -- (A4);
    \draw[very thick] (B1) -- (B4);
    \draw[very thick] (A2) -- (B4);
    \draw[very thick] (B3) -- (B4);

    \node[anchor=south east] at (A1) {$\kappa$-Galilei};
    \node[anchor=south east] at (A2) {Galilei};
    \node[anchor=south west] at (A3) {Newton - Hooke};
    \node[anchor=south west] at (A4) {$\kappa$-Newton - Hooke};
    \node[anchor=north west] at (B1) {$\kappa$-Poincar\'e};
    \node[anchor=north west] at (B2) {$\kappa$-(A) de Sitter};
    \node[anchor=south west] at (B3) {(A) de Sitter};
    \node[anchor=south west] at (B4) {Poincar\'e};

\draw[->,yshift=.2em,thin] (0.1, 1)--node[above]{$\Lambda$}(0.9, 1);
\draw[->,xshift=-.2em,thin] (0,0.9)--node[left]{$\frac 1 \kappa$}(0,0.2);
\draw[->,shift={(-.2em,-.2em)}, thin] (0,0)--node[anchor=north east]{$\frac 1 c $}(0.2,-0.2);

\end{tikzpicture}
\caption{The various algebras of relativistic symmetries emerging in the regimes set by  different combinations of the cosmological constant $\Lambda$, the speed of light $c$ and the quantum deformation parameter $\kappa$. The arrows point in the direction where the indicated parameter becomes nonzero. In addition to the ones showed in the previous picture, here we also see the classical deformation direction governed by the speed of light $c$, linking special-relativistic-like symmetries and Galilean-like symmetries.}
\label{fig2}
\end{center}
\end{figure}

\begin{figure}
\begin{center}
\begin{tikzpicture}[thick,scale=4]
    \coordinate (A1) at (0, 0);
    \coordinate (A2) at (0, 1);
    \coordinate (A3) at (1, 1);
    \coordinate (A4) at (1, 0);
    \coordinate (B1) at (0.3, -0.3);
    \coordinate (B2) at (1.3, -0.3);
    \coordinate (B3) at (1.3, 0.7);
    \coordinate (B4) at (0.3, 0.7);
    \coordinate (C1) at (0.6, -0.6);
    \coordinate (C2) at (1.6, -0.6);
    \coordinate (C3) at (1.6, 0.4);
    \coordinate (C4) at (0.6, 0.4);

    \draw[very thick] (A1) -- (A2) ;
    \draw[very thick] (A2) -- (A3);
    \draw[dashed] (A3) -- (A4);
    \draw[dashed] (A1) -- (A4);

    \draw[very thick] (A1) -- (B1);
    \draw[dashed] (B1) -- (B2);
    \draw[dashed] (B2) -- (B3);
    \draw[very thick] (B3) -- (A3);
    \draw[dashed] (B2) -- (A4);
    \draw[very thick] (B1) -- (B4);
    \draw[very thick] (A2) -- (B4);
    \draw[very thick] (B3) -- (B4);
    
    \draw[very thick] (B1) -- (C1);
    \draw[very thick] (C1) -- (C2);
    \draw[very thick] (C2) -- (C3);
    \draw[very thick] (C3) -- (B3);
    \draw[dashed] (C2) -- (B2);
    \draw[very thick] (C1) -- (C4);
    \draw[very thick] (B4) -- (C4);
    \draw[very thick] (C3) -- (C4);

    \node[anchor=south east] at (A1) {$\kappa$-Galilei};
    \node[anchor=south east] at (A2) {Galilei};
    \node[anchor=south west] at (A3) {Newton - Hooke};
    \node[anchor=south west] at (A4) {$\kappa$-Newton - Hooke};
    \node[anchor=south west] at (B1) {$\kappa$-Poincar\'e};
    \node[anchor=south west] at (B2) {$\kappa$-(A) de Sitter};
    \node[anchor=south west] at (B3) {(A) de Sitter};
    \node[anchor=south west] at (B4) {Poincar\'e};
    \node[anchor=north west] at (C1) {$\kappa$-Carroll};
    \node[anchor=north west] at (C2) {curved $\kappa$-Carroll};
    \node[anchor=south west] at (C3) {curved Carroll};
    \node[anchor=south west] at (C4) {Carroll};

\draw[->,yshift=.2em,thin] (0.1, 1)--node[above]{$\Lambda$}(0.9, 1);
\draw[->,xshift=-.2em,thin] (0,0.9)--node[left]{$\frac 1 \kappa$}(0,0.2);
\draw[->,shift={(-.2em,-.2em)}, thin] (0,0)--node[anchor=north east]{$\frac 1 c $}(0.2,-0.2);
\draw[->,shift={(-.2em,-.2em)}, thin] (0.5,-0.5)--node[anchor=north east]{$ c $}(0.3,-0.3);

\end{tikzpicture}
\caption{The various algebras of relativistic symmetries emerging in the regimes set by  different combinations of the cosmological constant $\Lambda$, the speed of light $c$ and the quantum deformation parameter $\kappa$. The arrows point in the direction where the indicated parameter becomes nonzero. In addition to the ones showed in the previous pictures, here we also see a new direction in which the classical deformation governed by the speed of light $c$ can work, linking special-relativistic-like symmetries and Carrollian-like symmetries.}
\label{fig3}
\end{center}
\end{figure}
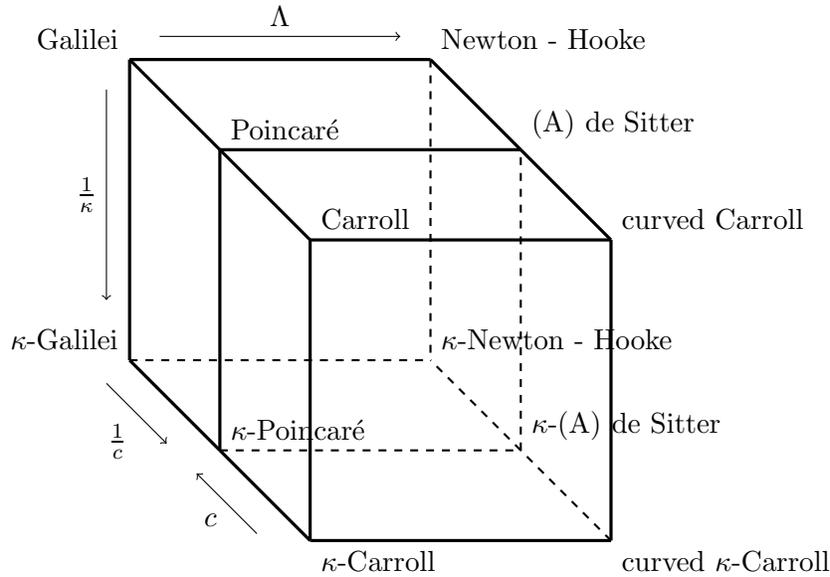

Very recent work analyzed yet another direction of classical deformation, this time governed by the speed of light $c$ (see figures \ref{fig2} and \ref{fig3}). The novel feature of this deformations with respect to the classical deformation governed by $\Lambda$ is that it can work in two different directions: starting from the Poincar\'e Lie algebra one can perform two kinds of contractions, one where $c^{-1}\rightarrow 0$ and one where $c\rightarrow 0$, which lead to the Galilei and Carroll Lie algebras and groups, respectively~\cite{LevyLeblondCarroll,BLL,Aldrovandi,Duval:2014uoa}. These two contractions can also be performed in the presence of the cosmological constant $\Lambda$ and of the quantum deformation parameter $\kappa$, as it was shown very recently \cite{BGGH2020kappanewtoncarroll}, thus providing us with a quite rich structure of possible  algebras of relativistic symmetries, shown in figure \ref{fig3}. We recall that Galilean symmetries with $\Lambda\neq 0$ are known in the literature as Newton-Hooke algebras~\cite{BLL}.

In this review, we survey the properties and relations of all of these algebras, emphasizing the different effects the three deformation parameters have and how they interact with one another. While the technical results on which we base our discussion have appeared in previous works, which are referenced to in the appropriate sections, this is the first time that a systematic picture of the properties and relations of these algebras is provided. 

The plan of this review is the following. In section \ref{sec:kP} we revisit the quantum deformation procedure turning the Poincar\'e Lie algebra into the $\kappa$-Poincar\'e Hopf algebra. In section \ref{sec:curvquant} we revisit the classical deformation procedure that turns the Poincar\'e algebra into the (Anti-)de Sitter algebra with non-vanishing cosmological constant and show how the quantum deformation procedure applies to the latter. The interplay between the effects of curvature and of quantum deformation are discussed. In section \ref{sec:curvlight} we perform the two classical contraction procedures governed by the speed of light, leading to the Galilean and Carrollian limits of  the classical (Anti-)de Sitter algebra. Here we discuss how the two classical deformations, governed by the speed of light and curvature, interact. Section \ref{sec:threeparams} looks at the full picture, where all of the three parameters are into play. The different features of the various algebras are revisited from the noncommutative spacetime point of view in section \ref{sec:ncst}. Final remarks are provided in section \ref{sec:conclusions}.

\section{The $\kappa$-Poincar\'e model}\label{sec:kP}

We start by briefly reviewing the classical  (3+1)-dimensional  Poincar\'e Lie algebra $\mathfrak{p}(3+1)$, using a language that will provide useful to discuss its quantum deformation. This algebra is defined by the commutation relations
\be
\begin{array}{lll}
[J_a,J_b]=\epsilon_{abc}J_c\,,& \quad [J_a,P_b]=\epsilon_{abc}P_c\,, &\quad
[J_a,K_b]=\epsilon_{abc}K_c\,, \\[2pt]
\displaystyle{
  [K_a,P_0]=P_a  }\,, &\quad\displaystyle{[K_a,P_b]=\delta_{ab} P_0}\,,    &\quad\displaystyle{[K_a,K_b]=-\epsilon_{abc} J_c}\,, 
\\[2pt][P_0,P_a]=0\, , &\quad   [P_a,P_b]=0\,, &\quad[P_0,J_a]=0\, ,
\end{array}
\label{pconm}
\ee
where in the so-called kinematical basis $\{P_0,P_a, K_a, J_a\}$  $(a=1,2,3)$ are the generators of time translations, space translations, boosts and rotations, respectively. Sum over repeated indices is assumed and for the moment the speed of light $c$ is set to $1$. As for any Lie algebra, the universal enveloping algebra $U(\mathfrak{p}(3+1))$ of the Poincar\'e algebra is a Hopf algebra endowed with a primitive (non-deformed) coproduct
\be
\Delta (X) = X \otimes 1 + 1 \otimes X, 
\qquad \forall X\in \mathfrak{p}(3+1)\, .
\label{coprim}
\ee
For the  generators of spacetime translations, this coproduct encodes algebraically the linear addition law for momenta that characterizes the usual special relativistic kinematics.

In this group-theoretical setting, Minkowski spacetime ${\mathbf
M}^{3+1}$ can be constructed from the Poincar\'e Lie group as the homogeneous space
\be
{\mathbf
M}^{3+1} \equiv  {\rm  ISO}(3,1)/{\rm  SO}(3,1)\, ,
\ee
where the isotropy subgroup is the Lorentz group SO(3,1). Explicitly, a 5-dimensional faithful representation $\rho$ for a generic element $X$ of the Poincar\'e Lie algebra is given by:
\begin{equation}
\label{eq:repG}
\rho(X)=   x^\alpha \rho(P_\alpha)  +  \xi^a \rho(K_a) +  \theta^a \rho(J_a) =\left(\begin{array}{ccccc}
0&0&0&0&0\cr 
x^0 &0&\xi^1&\xi^2&\xi^3\cr 
x^1 &\xi^1&0&-\theta^3&\theta^2\cr 
x^2 &\xi^2&\theta^3&0&-\theta^1\cr 
x^3 &\xi^3&-\theta^2&\theta^1&0
\end{array}\right) .
\end{equation}
If we parametrize an element $G$ of the Poincar\'e group ISO(3,1) in the form
\begin{align}
\begin{split}
\label{eq:Gm}
&G= \exp{x^0 \rho(P_0)} \exp{x^1 \rho(P_1)} \exp{x^2 \rho(P_2)} \exp{x^3 \rho(P_3)} \\
&\qquad\quad\times \exp{\xi^1 \rho(K_1)} \exp{\xi^2 \rho(K_2)} \exp{\xi^3 \rho(K_3)}
 \exp{\theta^1 \rho(J_1)} \exp{\theta^2 \rho(J_2)} \exp{\theta^3 \rho(J_3)} \, ,
\end{split}
\end{align}
the (3+1)-dimensional  Minkowski spacetime ${\mathbf M}^{3+1}$ can be constructed as a coset space (note that the Lorentz subgroup is located at the rightmost side in the exponentials above), whose points are labeled by the usual Minkowski coordinates $x^\alpha$ associated to translations. From a Hopf-algebraic point of view, this means that there is a pairing
\be
\langle  x^\alpha, P_\beta \rangle= \delta_\beta^\alpha \, .
\ee
between Poincar\'e translation generators and the  Minkowski coordinates $ x^\alpha$.

The representation theory of the Poincar\'e Lie algebra is characterized by its Casimir operators (see, for instance~\cite{WuKiTung}): the quadratic one 
\be
{\cal C}
= P_0^2-\>P^2  \, ,
\label{quadcas}
\ee
whose realization on momentum space gives rise to the energy-momentum dispersion relation, 
and the quartic one ${\cal W}$ constructed in terms of the components of the Pauli-Lubanski four vector in the form
\bea
&&  {\cal W}
=W^2_0-\>{W}^2 \, ,
\qquad\mbox{where}\label{PauliLubanski}\\[2pt]
&&  W_0= \>J\producto\>P 
\qquad
 W_a=-  J_a P_0+\epsilon_{abc} K_b P_c  \, .
 \nonumber
\eea

\subsection{The $\kappa$-Poincar\'e quantum algebra}

The $\kappa$-Poincar\'e algebra~\cite{ LRNT1991} (see also~\cite{GKMMK1992, LNR1992fieldtheory}) is a quantum Poincar\'e algebra, {\em i.e.} a Hopf algebra deformation (see~\cite{ChariPressley1994,Majidb}) of the Poincar\'e algebra in terms of a quantum deformation parameter $\kappa^{-1}$. The essential feature of quantum deformations is that, in general, the deformation affects  both the defining commutation rules of the algebra (which turn out to be nonlinear) and  the coproduct map (for which the linear rule of superposition of generators is broken). 

The deformed commutation rules and the deformed coproducts have to be compatible in the sense that the latter have to be a homomorphism map for the former. Moreover, quantum deformations are smooth  in the sense that in the vanishing deformation parameter limit  the quantum algebra reduces to the initial  Lie algebra. All these conditions restrict the number of possible inequivalent quantum deformations of a Lie algebra. For the Poincar\'e case, the classification of all its possible quantum deformations was presented in~\cite{Zakrzewski1997}, and the analogue classification in the quantum group setting was given in~\cite{PW1996}.

The $\kappa$-Poincar\'e algebra is a very specific Hopf algebra deformation of the Poincar\'e algebra which was obtained through quantum group contraction techniques~\cite{CGST1991heisemberg,CGST1992,BGHOS1995quasiorthogonal} from the so-called Drinfel'd-Jimbo quantum deformation of the (Anti)-de Sitter Lie algebra~\cite{Drinfeld1987icm,Jimbo1985}. 
Explicitly, its commutation rules are given by a non-deformed sector
\be
\begin{array}{lll}
[J_a,J_b]=\epsilon_{abc}J_c\,,& \quad [J_a,P_b]=\epsilon_{abc}P_c\,, &\quad
[J_a,K_b]=\epsilon_{abc}K_c\,, \\[2pt]
\displaystyle{
  [K_a,P_0]=P_a  }\,, &\quad\displaystyle{[K_a,K_b]=-\epsilon_{abc} J_c}\,,  &\quad
\\[2pt][P_0,P_a]=0\, , &\quad   [P_a,P_b]=0\,, &\quad[P_0,J_a]=0\, ,
\end{array}
\label{kconm1}
\ee
together with the following deformed commutators
\be
[K_a,P_b]=\delta_{ab} \left( \frac{\kappa}{2} \left(1-{\rm e}^{-2   P_0/\kappa} \right) + \frac{1}{2 \kappa}\, \textbf{P}^2 \right) -  \frac{1}{\kappa}\, P_a P_b \, .
\label{kconm2}
\ee
The deformed coproduct map for the $\kappa$-Poincar\'e algebra reads
\begin{align}
\begin{split}
&\Delta (P_0) = P_0 \otimes 1 + 1 \otimes P_0, \\
&\Delta (J_a) = J_a \otimes 1 + 1 \otimes J_a, \\
&\Delta (P_a) = P_a \otimes 1 +{\rm  e}^{-  P_0 /\kappa} \otimes P_a, \\
&\Delta (K_a) = K_a \otimes 1 + {\rm e}^{-  P_0 /\kappa} \otimes K_a +  \frac{1}{\kappa}\, \epsilon_{abc} P_b \otimes J_c\, .
\end{split}
\label{cokp}
\end{align}
We stress that the $\kappa^{-1}\to0$ limit of all these expressions leads to the non-deformed Hopf algebra structure of the Poincar\'e algebra.

It is also worth to emphasize that this is an ``essential" deformation in the sense that the theory of quantum universal enveloping algebras ensures that there does not exist any change of basis that transforms the deformed coproduct~\eqref{cokp} into the non-deformed one~\eqref{coprim}. On the other hand, it is possible to find a (nonlinear) change of basis transforming the deformed commutation rules~\eqref{kconm1}-\eqref{kconm2} into the non-deformed ones~\eqref{pconm}. As expected, such transformation to the so-called ``classical basis"~\cite{KowalskiGlikman:2002we} for $\kappa$-Poincar\'e   provides a (quite cumbersome) deformed coproduct, and shows that in order to prevent inconsistencies, all models defined through quantum deformations have to accommodate the full Hopf algebra structure (commutation rules + coproduct) as their underlying symmetry.

Some features of this quantum deformation of the Poincar\'e algebra deserve some attention. Firstly, the existence of deformed commutation rules~\eqref{kconm2} implies that Casimir operators have to be also $\kappa$-deformed. In particular, the deformed quadratic Casimir is found to be
\be
 {\cal C}_\kappa ={4}{\kappa^2}\sinh^2(P_0/2\kappa) - e^{P_0/\kappa}\>P^2 \, ,
 \label{casbicr}
\ee
and obviously its $\kappa\to\infty$ limit leads to~\eqref{quadcas}. When the corresponding momentum space representation of the  $\kappa$-Poincar\'e algebra is considered~\cite{KowalskiGlikman:2002we,GM2013relativekappa, AmelinoCamelia:2011nt}, this Casimir gives rise to a deformed dispersion relation, which is the cornerstone of the quantum gravity phenomenology of the  $\kappa$-Poincar\'e model (see~\cite{Amelino-Camelia2010symmetry} for a review on the role of $\kappa$-Poincar\'e in Doubly Special Relativity models). 

The deformation of the Pauli-Lubanski Casimir~\eqref{PauliLubanski} reads (see~\cite{BHMN2017kappa3+1} and references therein) 
\bea
&&\!\!\!\! \!\!\!\! \!\!\!\! 
 {\cal W}_\kappa
=\left(  \cosh(P_0/\kappa)-  \frac{1}{4\kappa^2} \, e^{P_0/\kappa}   \>P^2\right)W^2_{\kappa,0}-\>{W}_\kappa^2  \,  ,
\qquad \mbox{where}\nonumber\\ 
&&\!\!\!\! \!\!\!\! \!\!\!\! 
  W_{\kappa,0}=e^{P_0/(2\kappa)}\,  \>J\producto\>P ,
\qquad
 W_{\kappa,a}=-  J_a \, \kappa \sinh(P_0/\kappa)+e^{P_0/\kappa}\epsilon_{abc} \left( K_b + \frac{1}{2\kappa} \, \epsilon_{bkl}J_k P_l \right)P_c   \, .
\label{casdeformedb}
\eea
Similarly to what happens in the non-deformed Poincar\'e case, Casimir operators label the irreducible representations of the $\kappa$-Poincar\'e Hopf algebra. The spin zero representation was already given in~\cite{LRNT1991}, where the corresponding $\kappa$-Klein--Gordon equation was proposed. Irreducible representations for arbitrary spin were constructed in~\cite{GKMMK1992} for both the massive and the massless cases, and the $\kappa$-Dirac equation was introduced in~\cite{LNR1992fieldtheory,Giller:1993du,Nowicki:1992if}.

Secondly,  the deformed coproduct of the $\kappa$-Poincar\'e algebra provides a non-primitive addition law for momenta
\bea
\Delta ( P_0 ) \!\!\!&=&\!\!\!  P_0 \otimes 1+1 \otimes P_0, \nonumber\\
\Delta ( P_a )  \!\!\!&=&\!\!\!  P_a\otimes  1 +e^{-P_0/\kappa}  \otimes P_a \, ,
\label{hsub}
\eea 
which encodes in algebraic terms the nontrivial properties of the geometry of the associated momentum space. These expressions imply that the momentum sector of the  $\kappa$-Poincar\'e algebra is a Hopf subalgebra, since the coproducts of momenta generators depend only on themselves. As we will see in the following Section, this is no longer the case when the spacetime curvature $\Lambda$ is considered.  Finally, it is worth to mention that the Lorentz generators do not close a Hopf subalgebra, since the coproducts~\eqref{cokp} for the boost generators include translations. Quantum Poincar\'e and (A)dS algebras endowed with quantum Lorentz subgroup have been recently characterized in~\cite{Ballesteros:2021inq}.

\subsection{The $\kappa$-Poincar\'e Lie bialgebra and $\kappa$-Minkowski spacetime}

The ambiguity in the selection of the basis of the quantum algebra does not affect the Lie bialgebra structure $\delta$ associated to the $\kappa$-Poincar\'e  algebra. In fact, this is an object that characterizes any quantum deformation in a unique way since it does not depend on  changes of basis of the type
\be
X'=X'(P_0,P_a,J_a,K_a,\kappa) \qquad \mbox{with}\qquad \lim_{\kappa\to\infty}{X'}=X,
\qquad \mbox{for}\qquad X\equiv\{P_0,P_a,J_a,K_a\}.
\ee
Such Lie bialgebra structure is obtained by taking the skew-symmetric part of the first order in $1/\kappa$ of the deformed coproduct~\eqref{cokp}, and reads
 \begin{align}\label{kPc}
\begin{split}
& \delta(P_0) = \delta(J_a) = 0 ,\\
& \delta(P_a) = \frac{1}{\kappa} P_a \wedge P_0 ,\\
& \delta(K_1) = \frac{1}{\kappa} (K_1 \wedge P_0 + J_2 \wedge P_3 - J_3 \wedge P_2), \\
& \delta(K_2) = \frac{1}{\kappa}  (K_2 \wedge P_0 + J_3 \wedge P_1 - J_1 \wedge P_3), \\
& \delta(K_3) = \frac{1}{\kappa}  (K_3 \wedge P_0 + J_1 \wedge P_2 - J_2 \wedge P_1) .
\end{split}
\end{align}
This cocommutator map $\delta: \mathfrak{p}(3+1)\to \mathfrak{p}(3+1) \otimes \mathfrak{p}(3+1)$ is defined on the undeformed Poincar\'e algebra, and can be obtained from the classical $r$-matrix that characterizes the $\kappa$-deformation,
\be \label{kPr}
r=\frac{1}{\kappa} (K_1 \wedge P_1 + K_2 \wedge P_2 + K_3 \wedge P_3) \,,
\ee
through
$
\delta(X) = [1\otimes X + X \otimes 1, r] 
$,  where $r$ is a solution of the modified classical Yang-Baxter equation. From this perspective, the $r$-matrix is the `minimal' object that defines a given quantum deformation: from it, the first order deformation of the coproduct can be obtained, and the semiclassical counterpart of the associated quantum group (a Poisson-Lie group) is uniquely defined. 
In the approach here presented, Lie bialgebra structures are used as the defining objects for quantum deformations, and the type of interplay among all the parameters arising in them can be already studied at the Lie bialgebra level (in particular, the theory of quantum group contractions is based on the contraction theory for Lie bialgebras~\cite{BGHOS1995quasiorthogonal}). A detailed description of Lie bialgebras and their role in quantum group theory can be found in~\cite{ChariPressley1994}, and a complete presentation of kinematical Lie bialgebras has been given in~\cite{GH2021symmetry}. 

We also stress that the Hopf subalgebra structure of the momentum sector~\eqref{hsub} is reflected at the Lie bialgebra level in the form
\be
\delta(P_0) =  0 , \qquad
 \delta (P_1)=   \frac{1}{\kappa} \, P_1\wedge P_0   ,
 \qquad  \delta(P_2)=  \frac{1}{\kappa}\,  P_2\wedge P_0 . 
\ee
This sub-Lie bialgebra structure for the momentum sector can be dualized to give rise to the so-called $\kappa$-Minkowski Lie algebra
\be
[X^0, X^a]=-\frac{1}{\kappa}\,  X^a    , \qquad
 [X^a, X^b]=0.
 \label{kminkowski}
\ee
This algebra can be identified with the one defining the $\kappa$-Minkowski non-commutative spacetime~\cite{Maslanka1993,MR1994, Zakrzewski1994poincare, LRNT1991}. Moreover, 
the  $\kappa$-Poincar\'e momentum space can be constructed as an orbit of a certain linear action of the $\kappa$-Minkowski Lie group~\cite{KowalskiGlikman:2002ft,KowalskiGlikman:2003we,Kowalski-Glikman:2013rxa}.  Such an orbit turns out to be (a half of) the (3+1) de Sitter space with curvature $1/\kappa^2$, and the deformed dispersion relation of the model can be thought of as the distance to the origin in such curved momentum space~\cite{GM2013relativekappa}.

\subsection{Applications}

The  $\kappa$-Poincar\'e model and its associated quantum geometry has been extensively used in the literature in order to study different explicit models dealing with both mathematical and physical features of quantum geometry which are expected to arise at the Planck scale. Without aiming to be exhaustive, some of the facets of $\kappa$-Poincar\'e algebra and $\kappa$-Minkowski spacetime that have been analyzed in the literature are the following ones (see also references therein):

\begin{itemize}

\item Deformed dispersion relations and Doubly Special Relativity \cite{AM2000waves,KowalskiGlikman:2002jr,Amelino-Camelia:2008aez,BorowiecPachol2009jordanian,BP2010sigma,BGMP2010,ABP2017}, in particular the first paper associating deformed dispersion relations to $\kappa$-Poincar\'e/$\kappa$-Minkowski~\cite{AM2000waves} and the review~\cite{Amelino-Camelia:2008aez}.

\item $\kappa$-deformed models of Relative Locality~\cite{GM2013relativekappa,AmelinoCamelia:2011nt,Gubitosi:2019ymi,ALR2011speed,Carmona:2011wc,
Carmona:2012un,Carmona:2017cry,TrapanoSenzaCuore1,TrapanoSenzaCuore2,Gio}, see also the first papers defining the theory of Relative Locality,~\cite{ACFKS1,ACFKS2,Amelino-Camelia:2010wxx}.

\item There is an interesting string of works on the representation theory of $\kappa$-Minkowski commutation relations~\cite{Dabrowski:2010yk,Agostini:2002de,Agostini2007jmp,Lizzi:2018qaf,Carotenuto:2020rgn}.

\item Another aspect of interest is the  differential geometry of $\kappa$-Minkowski spacetime (and generalizations), and its relationship with the $\kappa$-Poincar\'e group and with star products, \cite{Sitarz1995plb,deAzcarraga:1995uw,Mercati:2011aa,Mercati:2010qhd,DS2013jncg,JMPS2015}.

\item There is a vast literature on how to construct classical (in the sense of $\hbar =0$) and quantum noncommutative field theories that are symmetric under $\kappa$-Poincar\'e group and are based on different versions of $\kappa$-Minkowski spacetime. A non-exhaustive list is~\cite{Kosinski1999,Kosinski2001,Agostini:2002yd,DJMTWW2003epjc,Kosinski:2003xx,Amelino-Camelia2007,AmelinoCamelia:2007rn,FKN2007plb,Arzano2007,Daszkiewicz:2007az,Freidel:2007hk,Arzano:2009ci,Gravityquantumspacetime2010,DJP2014sigma,Juric2015,Meljanac:2016jwk,Loret:2016jrg,Arzano:2017uuh,Ballesteros:2018ghw,JackFields,Poulain2018a,Poulain2018,Juric2018,Mercati:2018hlc,Mathieu2020,Arzano:2020jro}, and references therein.

\item A crucial issue is what limits to the spacetime localizability of observables does a  $\kappa$-deformed theory imply~\cite{Kosinski2001,Lizzi:2018qaf,FuzzyWorldlines, Amelino-Camelia:2013nza, Amelino-Camelia:2012vlk}. Related to this, is the possibility of deformations or fuzziness of light cones~\cite{MS2018plblightcone,Arzano2018}.

\item An important consequence of $\kappa$-deformed spacetime symmetries and noncommutative spacetimes is the emergence of a curvature of momentum space, and related deformations of phase space~\cite{KowalskiGlikman:2003we,Blaut:2003wg,GM2013relativekappa,LSW2015hopfalgebroids,Lizzi:2020tci,Carmona:2019fwf}.

\item Finally, a recent line of research led to the development of a $\kappa$-deformed noncommutative version of the spaces of worldlines~\cite{BGH2019worldlinesplb,FuzzyWorldlines}.

\end{itemize}


It is worth emphasizing that most of the above-mentioned techniques and models have been exclusively developed for the $\kappa$-Poincar\'e case. Therefore, the approach that we summarize in the following sections provides the basis for the generalization of all these results and models when the cosmological constant parameter $\Lambda$ is non-vanishing and/or for the Galilean and Carrollian limits when $c\to \infty$ and $c\to 0$, respectively.

\section{Interplay between curvature and  quantum effects}\label{sec:curvquant}

If one aims at studying the effects of quantum-deformed relativistic symmetries in a cosmological context (as is e.g. the case in studies of the propagation of signals from astrophysical sources \cite{Amelino-Camelia:2008aez, Addazi:2021xuf}), the most natural option consists in the generalization of the  $\kappa$-Poincar\'e model to allow for a nonvanishing cosmological constant $\Lambda$. This  leads to a quantum-deformed (Anti)-de Sitter (hereafter (A)dS) model. 

Works in (1+1) and (2+1) dimensions  already suggested that there is a nontrivial interplay between the quantum deformation and curvature. In particular, once the quantum deformation is taken into account the effects that are classically associated to spacetime curvature acquire a new energy-dependence. For example, the travel time of massless particles acquire an energy dependence that  depends on the curvature and the quantum deformation parameter in a nontrivial way~\cite{Marciano:2010gq,Amelino-Camelia:2012vzf, Rosati:2015pga, Barcaroli:2015eqe} (similar results are found in the context of twist deformations, see \cite{Aschieri:2020yft, ABP2017}).
While the phenomenology of the $\kappa$-(A)dS model in (3+1) dimensions still has to be worked out, preliminary studies   show that in this case the interplay between quantum deformation and curvature can be even more virulent, as we will discuss in this section.

Despite the fact that the $\kappa$-Poincar\'e algebra was initially obtained as the quantum group contraction associated to the flat $\Lambda\to 0$ limit of the quantum $so(3,2)$ algebra~\cite{LRNT1991,LNR1991realforms}, neither the relation among the generators of such $so(3,2)$ quantum deformation and the kinematical generators $\{P_0,P_a, K_a, J_a\}$ nor the explicit role played by the cosmological constant $\Lambda$ in the quantum case were explored. This lack of information prevented any physical interpretation, as well as the construction of the corresponding quantum (A)dS spacetimes in terms of local coordinates. This started to change recently, since  a series of  papers have filled this gap by constructing the fully explicit $\kappa$-(A)dS model~\cite{BHMN2017kappa3+1} and its associated noncommutative spacetime~\cite{BGH2019spacetime}. The main features of the former will be summarized in this section following the presentation of the  $\kappa$-Poincar\'e model given in the previous section, while the latter will be presented in section \ref{sec:ncst}. We stress that throughout this construction the curvature $\Lambda$ will be always made explicit as a ``classical" curvature parameter whose $\Lambda\to 0$ limit leads exactly to the  $\kappa$-Poincar\'e model.

\subsection{(Anti-)de Sitter symmetries as a classical deformation of Poincar\'e symmetries}
Before going to the quantum-deformed (A)dS model, we briefly show that the standard (A)dS algebra can be seen as a classical deformation of the Poincar\'e algebra. This is based on writing the
\adsw Lie  algebra in (3+1)D in the following manner
\be
\begin{array}{lll}
[J_a,J_b]=\epsilon_{abc}J_c\,,& \quad [J_a,P_b]=\epsilon_{abc}P_c\,, &\quad
[J_a,K_b]=\epsilon_{abc}K_c\,, \\[2pt]
\displaystyle{
  [K_a,P_0]=P_a  }\,, &\quad\displaystyle{[K_a,P_b]=\delta_{ab} P_0}\,,    &\quad\displaystyle{[K_a,K_b]=-\epsilon_{abc} J_c}\,, 
\\[2pt][P_0,P_a]=-{\Lambda} \, K_a\,, &\quad   [P_a,P_b]={\Lambda} \, \epsilon_{abc}J_c\,, &\quad[P_0,J_a]=0\, ,
\end{array}
\label{adsLie}
\ee
where $\Lambda$ is the cosmological constant parameter. This Lie algebra is just a $\Lambda$ deformation of~\eqref{pconm}, and the latter is obtained in the smooth $\Lambda\to 0$ limit of~\eqref{adsLie}. In this way, the three relativistic spacetimes with constant curvature are obtained as the following maximally symmetric homogeneous spaces:
\begin{itemize}
\item For $\Lambda<0$ we have the $SO(3,2)$ symmetry algebra and the AdS spacetime ${\mathbf
{AdS}}^{3+1}$ is obtained as the coset space ${\rm SO}(3,2)/{\rm  SO}(3,1)$.

\item For $ \Lambda>0$ we have the $SO(4,1)$ symmetry algebra that gives rise to the de Sitter spacetime ${\mathbf
{dS}}^{3+1} \equiv   {\rm  SO}(4,1)/{\rm SO}(3,1)$.

\item Finally, for $\Lambda=0$ we recover the Poincar\'e algebra, and Minkowski spacetime is ${\mathbf
M}^{3+1} \equiv  {\rm  ISO}(3,1)/{\rm  SO}(3,1)$.
\end{itemize}

This approach provides (A)dS Casimir operators as a $\Lambda$-deformation of Poincar\'e invariants. The quadratic one being
\be 
{\cal C}
= P_0^2-\>P^2  -{\Lambda} \left(   \>J^2-\>K^2\right) \, ,
\label{casqads}
\ee
and the quartic one (of Pauli-Lubanski type) reads
\bea
&&  {\cal W}
=W^2_0-\>{W}^2-{\Lambda} \left(\>J\producto\>K \right)^2 \, , \label{PLAdS}\\[2pt]
\mbox{where} &&  W_0= \>J\producto\>P 
\qquad
\mbox{and} \qquad 
 W_a=-  J_a P_0+\epsilon_{abc} K_b P_c \, .
\nonumber
\eea

Two main features of the \adsw Lie algebra~\eqref{adsLie} are worth to be emphasized. Firstly, that space-time translation generators do not commute when $\Lambda\neq 0$:
\be
[P_0,P_a]=- {\Lambda} \, K_a\,, \quad   [P_a,P_b]= {\Lambda} \, \epsilon_{abc}J_c\, ,
\ee 
and therefore the translation sector does not define a Lie subalgebra. This reflects the fact that the (A)dS spacetimes are curved spaces, since spacetime translations are generators of geodesic motions on them.

Secondly, when $\Lambda\neq 0$ the following automorphism interchanges the role of $P_a$ and $K_a$ (see~\cite{Ballesteros:2017pdw}):
\be
\tilde P_0=P_0,\qquad \tilde P_a= \sqrt {-\Lambda}\, K_a,\qquad \tilde K_a=-\frac{1}{\sqrt{-\Lambda}}\, P_a,\qquad \tilde J_a= J_a .
\label{autom}
\ee
In this sense, translations and boosts play an algebraically equivalent role, albeit their physical meaning is indeed different. As we will see, this property will be essential in order to understand some of the features of the $\kappa$-(A)dS model.

\subsection{The $\kappa$-(A)dS model in (3+1) dimensions} 

We recall that the (2+1) dimensional $\kappa$-(A)dS algebra and deformed Casimir operators was already presented in~\cite{BHOS1994global}. The very same quantum algebra was later rediscovered in~\cite{ASS2004} as the algebra containing the cosmological constant that was proposed as a symmetry for the low energy limit of a quantum theory of gravity (see also~\cite{JackkdS} for a more recent approach). The classical $r$-matrix generating such a (2+1) quantum (A)dS deformation is
\be
r=\frac{1}{\kappa}( K_1 \wedge P_1 + K_2 \wedge P_2 )\, .
\label{r21}
\ee
Surprisingly enough, the cosmological constant parameter $\Lambda$ is absent in this $r$-matrix, which therefore coincides with its Poincar\'e limit. Nevertheless, the full quantum algebra does contain $\Lambda$ explicitly.  

The quest for the generalization of~\eqref{r21} to the (3+1)-dimensional case was recently solved in~\cite{BHMN2017kappa3+1}, and the unique (modulo automorphisms)  solution is
\be
r_\L=\frac{1}{\kappa}( K_1 \wedge P_1 + K_2 \wedge P_2 + K_3 \wedge P_3 + \eta  J_1 \wedge  J_2)\, ,
\label{r31}
\ee
where from now on we will use the parameter $ {\eta^2:=-\Lambda}$. This is the unique skewsymmetric $r$-matrix for the (A)dS algebra fulfilling two conditions: its $\Lambda\to 0$ limit of~\eqref{r31} is the $\kappa$-Poincar\'e $r$-matrix (this guarantees the appropriate flat limit of the model), and the cocommutator of the $P_0$ generator is primitive $\delta(P_0)=0$ (this enables in the curved case the interpretation of  $\kappa$ as a mass). 

From the $r$-matrix~\eqref{r21} the following $\kappa$-(A)dS cocommutator map is obtained
\begin{align}
\begin{split}
& \delta(P_0)=\delta(J_3)= 0, \qquad \delta(J_1)=\frac \r\kappa J_1 \wedge J_3, \qquad \delta(J_2)= \frac \r\kappa   J_2 \wedge J_3 ,\\
& \delta(P_1)= \frac{1}{\kappa} (P_1 \wedge P_0 - \r P_3 \wedge J_1 - \r^2 K_2 \wedge J_3 + \r^2 K_3 \wedge J_2) ,\\
& \delta(P_2)= \frac{1}{\kappa} (P_2 \wedge P_0 - \r P_3 \wedge J_2 + \r^2 K_1 \wedge J_3 - \r^2 K_3 \wedge J_1), \\
& \delta(P_3)= \frac{1}{\kappa} (P_3 \wedge P_0 + \r P_1 \wedge J_1 + \r P_2 \wedge J_2 - \r^2 K_1 \wedge J_2 + \r^2 K_2 \wedge J_1), \\
& \delta(K_1)= \frac{1}{\kappa} (K_1  \wedge P_0  + P_2 \wedge J_3 - P_3 \wedge J_2 - \r K_3 \wedge J_1) ,\\
& \delta(K_2)= \frac{1}{\kappa} ( K_2 \wedge P_0  - P_1 \wedge J_3 + P_3 \wedge J_1 - \r K_3 \wedge J_2) ,\\
& \delta(K_3)= \frac{1}{\kappa} ( K_3 \wedge P_0  + P_1 \wedge J_2 - P_2 \wedge J_1 + \r K_1 \wedge J_1 + \r K_2 \wedge J_2).
\end{split}
\label{cocoads}
\end{align} 
When comparing these expressions with the ones that hold for   $\kappa$-Poincar\'e  (which are recovered in the $\r\to 0$ limit), several features of the new model arise, which are not present in the $\kappa$-Poincar\'e  nor in the classical (A)dS limit, thus being due genuinely to the interplay between the two deformations. The most striking feature is that
the $so(3)$ subalgebra generated by rotations $J_a$ is  deformed, with a deformation governed by the ratio ${\eta}/{\kappa}$. Therefore, the deformation of space isotropy has to be expected as a direct consequence of the interplay between curvature and quantum effects. Moreover,  the cocommutator for the translations sector does no longer define a sub-Lie bialgebra structure, and involves the Lorentz sector. Related to this, the expressions for $\delta(P_i)$ and $\delta(K_i)$ can be interchanged under the automorphism~\eqref{autom}. Therefore, deformed space translations and boosts are expected to play similar algebraic roles within the $\kappa$-(A)dS model.

We recall that the cocommutator~\eqref{cocoads} provides the first order in the quantum deformation. In~\cite{BHMN2017kappa3+1}, by making use of a Poisson version of the so-called ``quantum duality principle" presented in~\cite{BM2013dual}, full expressions for the (Poisson) analogue of the full $\kappa$-(A)dS model were explicitly obtained. Here we recall only some of them in order to  illustrate the previous remarks. In particular, the rotations sector is deformed into a quantum $so(3)$ algebra with deformation parameter given by $ \eta/\kappa= \sqrt{-\Lambda}/\kappa$:
\bea
 \Delta  ( J_3  )  \!\!\!&=&\!\!\!    J_3 \otimes 1 +1 \otimes J_3 , \nonumber\\
\Delta  ( J_1  ) \!\!\!&=&\!\!\!     J_1 \otimes e^{\frac{\eta}{\kappa} J_3} +1 \otimes J_1 ,\qquad \Delta ( J_2 ) = J_2 \otimes e^{\frac{\eta}{\kappa}J_3}+1 \otimes J_2  , 
\label{deformedso3}
\eea
and whose deformed brackets read
\be
\begin{array}{lll}
\multicolumn{3}{l}
{\displaystyle {\left\{ J_1,J_2 \right\}= \frac{e^{2\frac{\eta}{\kappa} J_3}-1}{2 {\eta}/{\kappa} } - \frac{{\eta}}{2 \kappa} \left(J_1^2+J_2^2\right)  ,\qquad
\left\{ J_1,J_3 \right\}=-J_2 ,\qquad
\left\{ J_2,J_3 \right\}=J_1 \,  . } }
 \end{array}
\label{be}
\ee

The coproduct for the translations sector, that in principle would provide the deformed composition law for momenta in the corresponding DSR model, as seen for the $\kappa$-Poincar\'e case in the previous section,  reads
\bea
\Delta ( P_0 ) \!\!\!&=&\!\!\!  P_0 \otimes 1+1 \otimes P_0   , \nonumber \\
\Delta ( P_1 )  \!\!\!&=&\!\!\!  P_1\otimes   \cosh (\eta J_3 /\kappa) +e^{-P_0/\kappa}  \otimes P_1 
-  {\eta} K_2 \otimes     { \sinh (\eta J_3 /\kappa) }  \nonumber \\
&&- \frac{\eta}{\kappa}P_3  \otimes J_1 + \frac{\eta^2}{\kappa} K_3 \otimes J_2  + \frac{\eta^2}{\kappa^2} \left( {\eta }K_1-P_2 \right)  \otimes J_1 J_2 e^{-\frac{\eta}{\kappa}J_3}   \nonumber\\
&&-  \frac{\eta^2}{\kappa^2}  \left(  {\eta}K_2+P_1   \right)  \otimes      \left( J_1^2-J_2^2 \right)    e^{-\frac{\eta}{\kappa}J_3}   , \nonumber \\
\Delta ( P_2 ) \!\!\!&=&\!\!\!  P_2\otimes  \cosh( \eta J_3 /\kappa) +e^{-P_0/\kappa}  \otimes P_2
+  \eta K_1 \otimes    { \sinh (\eta J_3 /\kappa) }  \nonumber \\
&&- \frac{\eta}{\kappa} P_3  \otimes J_2 - \frac{\eta^2}{\kappa} K_3 \otimes J_1  - \frac{\eta^2}{\kappa^2} \left( \eta K_2+P_1 \right)  \otimes J_1 J_2 e^{-\frac{\eta}{\kappa}J_3}    \\
&& - \frac 12 \frac{\eta^2}{\kappa^2}    \left(    \eta K_1- P_2 \right)  \otimes     \left( J_1^2-J_2^2 \right)       e^{-\frac{\eta}{\kappa}J_3}  , \nonumber \\
\Delta ( P_3 )  \!\!\!&=&\!\!\!  P_3  \otimes 1 +e^{-P_0/\kappa}  \otimes P_3  + \frac{1}{\kappa}  \left(  \eta^2  K_2+ \eta  P_1 \right)  \otimes J_1  e^{-\frac{\eta}{\kappa}J_3}  \nonumber \\
&& -\frac{1}{\kappa}     \left(  \eta^2  K_1- \eta P_2 \right) \otimes  J_2 e^{-\frac{\eta}{\kappa}J_3}  \, . \nonumber 
\eea
As we anticipated from the Lie bialgebra structure, the deformed composition law for momenta involves the full Lorentz sector, which indicates that the construction of the  associated momentum needs to include the Lorentz sector as well~\cite{BGGHplb2017,Ballesteros:2017pdw}.
Moreover, the corresponding deformed brackets show that momenta are both non-commuting (due to $\eta\neq 0$) and quantum deformed:
  \begin{eqnarray}
&&\!\!\!\! \!\!\!\! \left\{ P_1, P_2 \right\}=- \eta^2\, \frac{  \sinh(2 \frac{\eta}{\kappa}J_3)}{2{\eta}/{\kappa} }- \frac{\eta}{2 \kappa}  \left( 2 P_3^2+ {\eta^2} (J_1^2+J_2^2)  \right)- \frac{\eta^5}{4 \kappa^3} e^{-2\frac{\eta}{\kappa}J_3} \left(J_1^2+J_2^2 \right)^2  \, , \nonumber \\
&&\!\!\!\! \!\!\!\! \left\{ P_1, P_3 \right\}=\frac12 {\eta^2} J_2 \left( 1+e^{-2\frac{\eta}{\kappa}J_3} \left[1+   \frac{\eta^2}{\kappa^2}   \left( J_1^2 +J_2^2 \right) \right]  \right) + \frac{\eta}{\kappa}  P_2 P_3  \, , \\
&& \!\!\!\! \!\!\!\! \left\{ P_2, P_3 \right\}=-\frac12 {\eta^2} J_1 \left( 1+e^{-2\frac{\eta}{\kappa}J_3} \left[1+  \frac{\eta^2}{\kappa^2}   \left( J_1^2 +J_2^2 \right) \right]  \right)  - \frac{\eta}{\kappa}   P_1 P_3\,. \nonumber
\end{eqnarray}
Note also here the complicated interplay between curvature and quantum effects arising in the quantum deformation, which is expressed through products of different powers of $1/\kappa$ and of the cosmological constant parameter $\eta$. Nevertheless, we stress that we have an all-order model at hand, with which all types of DSR predictions could be in principle computed.

Finally, we recall the (Poisson) counterpart of the second-order Casimir
\bea
{\cal C}&=&2{\kappa^2}\left[ \cosh (P_0/\kappa)\cosh(\frac{\eta}{\kappa} J_3)-1 \right]+ {\eta^2} \cosh(P_0/\kappa) (J_1^2+
J_2^2) e^{-\frac{\eta}{\kappa}J_3} \nonumber\\
&& -e^{P_0/\kappa} \left( \mathbf{P}^2 + {\eta^2} \mathbf{K}^2 \right)   \left[ \cosh(\frac{\eta}{\kappa}J_3)+ \frac {\eta^2 }{2\kappa^2}  (J_1^2+J_2^2)  e^{-\frac{\eta}{\kappa}J_3} \right]\nonumber\\
&&+2 {\eta^2} e^{P_0/\kappa}  \left[ \frac{\sinh(\frac{\eta}{\kappa}J_3)}{\eta}\R_3+ \frac{1}{\kappa} \left( J_1\R_1 +J_2 \R_2+  \frac {\eta}{2\kappa} (J_1^2+J_2^2) \R_3 \right)  e^{-\frac{\eta}{\kappa} J_3} \right],
\eea
where $\R_a=\epsilon_{abc} K_b P_c $. As expected, in the $\kappa\to \infty$ limit we obtain~\eqref{casqads}, and in the $\eta\to 0$ limit, we obtain the $\kappa$-Poincar\'e quantum Casimir in the bicrossproduct basis~\eqref{casbicr}.
The Poisson version of the $\kappa$-(A)dS analogue of the Pauli-Lubanski fourth order Casimir~\eqref{PLAdS} was also presented in~\cite{BHMN2017kappa3+1},
and the study of the representation theory of the $\kappa$-(A)dS Hopf algebra is still an open problem.

\section{Interplay between curvature and the speed of light} \label{sec:curvlight}

So far, the speed of light parameter has been set to $c=1$. Therefore, in order to unveil the coupling between $\Lambda$ and $c$, the latter parameter has to be explicitly included in the formalism.  At the classical level it is well-known~\cite{LevyLeblondCarroll,BLL,Duval:2014uoa} that this gives rise to two possible limits: the so-called ``non-relativistic" or ``Galilean" limit $c\to \infty$ and the ``ultra-relativistic" or `Carrollian" limit $c\to 0$. A complete study of the metrics and foliations for classical Galilei and Carroll spaces (also in the curved cases with$\Lambda\neq 0$) can be found in the literature (see, for instance,~\cite{SnyderG} and references therein).

\subsection{The Galilean limit of (A)dS}

The Galilean limit corresponds to taking small velocities compared to the speed of light. In this limit the light-cone opens along $t=0$, producing a space with absolute time.

The interplay between the contraction procedure and curvature can be studied by looking at the contraction of the (A)dS spacetime and its algebra of symmetries. This is obtained via an In\"on\"u--Wigner contraction procedure, induced by the algebra automorphism $\mathcal P(P_{0}, P_{a},K_{a},J_{a})=(P_{0}, -P_{a},-K_{a},J_{a})$ (speed-space contraction), see for example \cite{SnyderG}. Upon the rescaling 
\be
P_{a}\rightarrow \frac{P_{a}}{c}\, , \qquad K_{a}\rightarrow \frac{K_{a}}{c} \, , \label{eq:GalileiRescaling}
\ee
one finds that when $c\rightarrow \infty$ the following commutators of the (A)dS algebra are modified:
\be
\begin{array}{ll}
\left[K_{a},P_{b}\right]=\delta_{ab}\frac{1}{c^{2}}P_{0}&\rightarrow \left[K_{a},P_{b}\right]=0\\
\left[K_{a},K_{b}\right]=-\epsilon_{abc}\frac{1}{c^{2}}J_{c}&\rightarrow \left[K_{a},K_{b}\right]=0\\
\left[P_{a},P_{b}\right]=\Lambda \epsilon_{abc}\frac{1}{c^{2}}J_{c}&\rightarrow \left[P_{a},P_{b}\right]=0\,,
\end{array}
\ee
and the Casimir reduces to
\be
\mathcal C=\>P^2-\Lambda \>K^2\,.
\ee

We note that the presence of curvature does not affect the appearance of an absolute time in the Galilean limit, since the commutator between boosts and time translation vanishes. However, while in the flat $\Lambda\to 0$ limit the translation sector in unaffected by the Galilei contraction, when curvature is present one still gets `flat' spatial slices in the Galilei limit, since the commutator between spatial translations vanishes (see~\cite{SnyderG} for details).

\subsection{The Carroll limit of (A)dS}
The Carroll limit corresponds to taking large space intervals. It is relevant in the strong gravity regime and close to the black hole horizon (see~\cite{Bergshoeff:2017btm} and references therein). In contrast to the Galilean limit, in this case the light-cone closes along the $t$ direction.

As done in the Galilean case, we look at the contraction of the (A)dS spacetime and its algebra of symmetries. This is obtained via an In\"on\"u--Wigner contraction procedure, induced by the algebra automorphism $\mathcal T(P_{0}, P_{a},K_{a},J_{a})=(-P_{0}, P_{a},-K_{a},J_{a})$ (speed-time contraction), see for example \cite{SnyderG}. Upon the rescaling 
\be
P_{0}\rightarrow c P_{0}\, , \qquad K_{a}\rightarrow c K_{a}\, ,  \label{eq:CarrollRescaling}
\ee
one finds that when $c\rightarrow 0$ the following commutators of the (A)dS algebra are modified:
\be
\begin{array}{ll}
\left[K_{a},K_{b}\right]=-\epsilon_{abc}c^{2} J_{c}&\rightarrow \left[K_{a},K_{b}\right]=0\\
\left[K_{a},P_{0}\right]=c^{2} P_{a}&\rightarrow \left[K_{a},P_{0}\right]=0\,,
\end{array}
\ee
and the Casimir reduces to
\be
\mathcal C=  P_{0}^{2}+\Lambda \>K^{2}\,.
\ee

Similarly to the Galilean case, the most relevant feature of the Carrollian relativity, namely that of having an absolute space, is preserved in presence of curvature. Moreover, the nocommutativity of translations, caused by spacetime curvature, is not affected in the Carrollian limit, as opposed to what happens in the Galilean case. A summary of the different effects that the non-relativistic and the ultra-relativistic limits  have on the symmetries of a given spacetime with and without curvature is presented in Table \ref{table1}.

\begin{table}[h]
{\footnotesize
\caption{\small  
Table with the summary of the interplay between curvature and the speed of light parameter as seen in the (A)dS algebra and its Galilean and Carrollian limits. Horizontal lines indicate that the commutator is the same for the three cases.}
\label{table1}
  \begin{center}
\noindent
\begin{tabular}{l c c c  }
\hline
\\[-0.2cm] 
&Galilean limit  &  (A)dS &  Carrollian limit  \\[0.2cm]
\hline
\\[-0.2cm]
$[J_{a},J_{b}]$&\multicolumn{3}{c}{\rule[2pt]{3cm}{0.2pt}  $\epsilon_{abc} J_{c}$ \rule[2pt]{3cm}{0.2pt}} \\[0.2cm]
$[J_{a},P_{b}]$&\multicolumn{3}{c}{\rule[2pt]{3cm}{0.2pt}  $\epsilon_{abc} P_{c}$ \rule[2pt]{3cm}{0.2pt}}\\[0.2cm]
$[J_{a},K_{b}]$&\multicolumn{3}{c}{\rule[2pt]{3cm}{0.2pt}  $\epsilon_{abc} K_{c}$ \rule[2pt]{3cm}{0.2pt}}\\[0.2cm]
$[J_{a},P_{0}]$&\multicolumn{3}{c}{\rule[2pt]{3.4cm}{0.2pt} $0$ \rule[2pt]{3.4cm}{0.2pt}}\\[0.2cm]
\hline
\\[-0.2cm] 
$[K_{a},K_{b}]$&$0$&$-\epsilon_{abc} J_{c}$&$0$ \\[0.2cm]
$[K_{a},P_{b}]$&$0$&$\delta_{ab} P_{0}$ & $\delta_{ab} P_{0}$\\[0.2cm]
$[K_{a},P_{0}]$&$P_{a}$&$P_{a}$&$0$\\[0.2cm]
\hline
\\[-0.2cm]
$[P_{a},P_{b}]$&$0$&$\Lambda \epsilon_{abc}J_{c}$ & $\Lambda \epsilon_{abc}J_{c}$\\[0.2cm]
$[P_{a},P_{0}]$&$\Lambda K_{a}$&$\Lambda K_{a}$&$\Lambda K_{a}$\\[0.2cm]
\hline
\end{tabular}
 \end{center}
}
 \end{table}


\newpage

\section{Interplay of the three parameters: curvature, speed of light and quantum deformation} \label{sec:threeparams}

\subsection{Zero curvature case: Galilei and Carroll contraction of $\kappa$-Poincar\'e}

In order to study the Galilei and Carroll limits of the $\kappa$-Poincar\'e algebra, we would like to perform a contraction similar to the one used in the non-quantum case of the previous section. However, as was discussed in detail in \cite{BGGH2020kappanewtoncarroll}, the contraction procedure of a quantum algebra (Lie bialgebra contraction) might require a rescaling of the quantum deformation parameter along with the generators in order to obtain meaningful structures.

In general one can perform two kinds of contractions, either working at the level of the r-matrix (this is a ``coboundary" contraction), or working directly at the level of the co-commutators (this is the so-called ``fundamental" contraction)~\cite{BGHOS1995quasiorthogonal,GH2021symmetry}.
As it was shown in~\cite{BGGH2020kappanewtoncarroll}, this distinction is especially relevant in the case of the Galilean limit of $\kappa$-Poincar\'e, where the two procedures are nonequivalent. In fact, after the rescaling \eqref{eq:GalileiRescaling}, the $\kappa$-Poincar\'e r-matrix \eqref{kPr} reads:
\be
r=\frac{c^{2}}{\kappa}\left( K_{1} \wedge P_{1}+ K_{2} \wedge P_{2}+K_{3} \wedge P_{3} \right)\,.
\ee
This is well-behaved in the $c\rightarrow\infty$ limit if also the quantum parameter is rescaled as $\kappa\rightarrow \kappa/c^{2}$. However the resulting r-matrix
\be
r=\frac{1}{\kappa}\left( K_{1} \wedge P_{1}+ K_{2} \wedge P_{2}+K_{3} \wedge P_{3} \right)\, ,
\ee
produces trivial cocommutators, $\delta(X)=0$ for all generators $X$ of the algebra. So the coboundary contraction of the $\kappa$-Poincar\'e algebra produces the classical Galilei algebra.
On the other hand, working directly at the level of the cocommutators \eqref{kPc}, one can easily see that they are invariant under the rescaling \eqref{eq:GalileiRescaling}, so that the  $c\rightarrow\infty$ limit is well-defined without need to rescale the quantum deformation parameter. The resulting $\kappa$-Galilei algebra contains the following modified commutators with respect to the $\kappa$-Poincar\'e algebra, which corresponds to the left column, and in which the automorphism~\eqref{eq:GalileiRescaling} has been applied:
\be
\begin{array}{ll}
\left[K_{a},P_{b}\right]=\frac{\delta_{ab}}{c^{2}}\left[ \frac{\kappa}{2}\left(1-e^{-2P_{0}/\kappa}\right)+c^{2}\frac{\vec P^{2}}{2\kappa}\right]-\frac{P_{a}P_{b}}{\kappa}&\rightarrow \left[K_{a},P_{b}\right]=\delta_{ab} \frac{\vec P^{2}}{2\kappa} -\frac{P_{a}P_{b}}{\kappa}\\
\left[K_{a},K_{b}\right]=-\frac{\epsilon_{abc}}{c^{2}}J_{c}&\rightarrow \left[K_{a},K_{b}\right]=0\,,
\end{array}
\ee
while the coproducts are unmodified with respect to the $\kappa$-Poincar\'e case.

When performing the Carrollian limit of the $\kappa$-Poincar\'e algebra, one finds that the two procedures outlined above give equivalent results. The rescaled r-matrix reads
\be
r=\frac{1}{c \kappa}\left( K_{1} \wedge P_{1}+ K_{2} \wedge P_{2}+K_{3} \wedge P_{3} \right)\,,
\ee
which is well-behaved in the $c\rightarrow 0$ if the quantum deformation parameter is rescaled as $\kappa\rightarrow c\kappa$.
Then the r-matrix   reads
\be
r=\frac{1}{\kappa}\left( K_{1} \wedge P_{1}+ K_{2} \wedge P_{2}+K_{3} \wedge P_{3} \right)\,,
\ee
and produces non-trivial co-commutators:
\be
\begin{array}{ll}
\delta(P_{0})&=\delta(J_{a})=0\, , \\
\delta(P_{a})&=\frac{1}{\kappa} P_{a}\wedge P_{0}\, , \\
\delta(K_{a})&=\frac{1}{\kappa} K_{a}\wedge P_{0}\,.
\end{array}
\ee
The resulting $\kappa$-Carroll algebra contains the following modified commutators with respect to the $\kappa$-Poincar\'e algebra:
\be
\begin{array}{ll}
\left[K_{a},P_{0}\right]=P_{a} c^{2}&\rightarrow \left[K_{a},P_{0}\right]=0\\
\left[K_{a},P_{b}\right]=c \delta_{ab}\left[ \frac{\kappa}{2 c}\left(1-e^{-2P_{0}/\kappa}\right)+\frac{\vec P^{2} c}{2\kappa}\right]-c \frac{P_{a}P_{b}}{\kappa}&\rightarrow \left[K_{a},P_{b}\right]=\delta_{ab} \frac{\kappa}{2}\left(1-e^{-2P_{0}/\kappa}\right)\\
\left[K_{a},K_{b}\right]=-c^{2}\epsilon_{abc}J_{c}&\rightarrow \left[K_{a},K_{b}\right]=0\,,
\end{array}
\ee
while, again, the coproducts are unmodified.

We see that in both the Galilean and Carrollian limits the commutator between boosts generators vanishes, as in the classical case. A relevant difference between the two limits is that, while in the Carrollian limit the presence of the quantum deformation does not spoil the appearance of an absolute space (signaled by the vanishing of the commutator between boosts and time translations), in the Galilean limit the mixing between time and space induced by the quantum deformation prevents the emergence of an absolute time, since the commutators between boosts and spatial translations remain non-vanishing in the transition from the $\kappa$-Poincar\'e to the $\kappa$-Galilei symmetries. These properties are summarized in Table \ref{table2}.

\begin{table}[h]
{\footnotesize
\caption{\small  
Summary of the properties of the  $\kappa$-Galilei,  $\kappa$-Poincar\'e and  $\kappa$-Carroll algebras.}
\label{table2}
  \begin{center}
\noindent
\begin{tabular}{l c c c  }
\hline
\\[-0.2cm] 
&$\kappa$-Galilei  & $\kappa$-Poincar\'e &  $\kappa$-Carroll  \\[0.2cm]
\hline
\\[-0.2cm]
$[J_{a},J_{b}]$&\multicolumn{3}{c}{\rule[2pt]{3cm}{0.2pt}  $\epsilon_{abc} J_{c}$ \rule[2pt]{3cm}{0.2pt}} \\[0.2cm]
$[J_{a},P_{b}]$&\multicolumn{3}{c}{\rule[2pt]{3cm}{0.2pt}  $\epsilon_{abc} P_{c}$ \rule[2pt]{3cm}{0.2pt}}\\[0.2cm]
$[J_{a},K_{b}]$&\multicolumn{3}{c}{\rule[2pt]{3cm}{0.2pt}  $\epsilon_{abc} K_{c}$ \rule[2pt]{3cm}{0.2pt}}\\[0.2cm]
$[J_{a},P_{0}]$&\multicolumn{3}{c}{\rule[2pt]{3.4cm}{0.2pt} $0$ \rule[2pt]{3.4cm}{0.2pt}}\\[0.2cm]
\hline
\\[-0.2cm] 
$[K_{a},K_{b}]$&$0$&$-\epsilon_{abc} J_{c}$&$0$ \\[0.2cm]
$[K_{a},P_{b}]$&$\frac{\delta_{ab}}{2\kappa}\vec P^{2}-\frac{P_{a}P_{b}}{\kappa}$&$\delta_{ab}\left[\frac{\kappa}{2}\left(1-e^{-2P_{0}/\kappa}\right)+\frac{\vec P^{2}}{2\kappa}\right]-\frac{P_{a}P_{b}}{\kappa}$ & $\delta_{ab}\frac{\kappa}{2}\left(1-e^{-2P_{0}/\kappa}\right)$\\[0.2cm]
$[K_{a},P_{0}]$&$P_{a}$&$P_{a}$&$0$\\[0.2cm]
\hline
\\[-0.2cm]
$[P_{a},P_{b}]$&\multicolumn{3}{c}{\rule[2pt]{3cm}{0.2pt}  $0$ \rule[2pt]{3cm}{0.2pt}}\\[0.2cm]
$[P_{a},P_{0}]$&\multicolumn{3}{c}{\rule[2pt]{3cm}{0.2pt}  $0$ \rule[2pt]{3cm}{0.2pt}}\\[0.2cm]
\hline
\end{tabular}
 \end{center}
}
 \end{table}

\subsection{With curvature: Galilei and Carroll contraction of $\kappa$-(A)dS }\label{sub:kAdScontractions}

Here we study the interplay of all of the three parameters that govern different kinds of deformations of special relativity: the speed of light, the cosmological constant and the quantum deformation parameter.

In order to do so, we look at the Galilean and Carrollian contraction of the $\kappa$-(A)dS algebra. This is done by following the same procedure discussed in the previous subsection for the $\Lambda=0$ case. 
The detailed formulas can be found in \cite{BGGH2020kappanewtoncarroll} and are schematically represented in Table \ref{table3}. Here we discuss the points that are particularly relevant. We noticed in section \ref{sec:curvquant} that an important effect of the interplay between curvature and quantum deformation is that the rotation sector gets deformed. The Galilean contraction does not spoil this feature, while the Carrollian contraction restores standard isotropy. As already observed in the $\Lambda=0$ case, the  mixing between time and space due to the quantum deformation prevents the emergence of an absolute time in the Galilean limit, and the presence of curvature does not affect this result. Finally, one can see effects that are only relevant when all of the three parameters enter in the deformation of the Poincar\'e algebra: in the Galilean limit, when the curvature is non-zero, the commutator between boosts does not vanish, and is proportional to $\sqrt{\Lambda}/\kappa$. In general, the Carrollian limit seems to be a ``milder'' deformation, since it is isotropic, preserve the absoluteness of space and the vanishing commutators between boosts.

\begin{table}[h]
{\footnotesize
\caption{\small  
Summary of the properties of curved $\kappa$-Galilei,  $\kappa$-(A)dS and  curved $\kappa$-Carroll
}
\label{table3}
  \begin{center}
\noindent
\begin{tabular}{l c c c  }
\hline
\\[-0.2cm] 
&(curved) $\kappa$-Galilei & $\kappa$-(A)dS &  (curved) $\kappa$-Carroll  \\[0.2cm]
\hline
\\[-0.2cm]
$[J_{a},J_{b}]$&\rule[-6pt]{0.2pt}{20pt}   &\rule[-6pt]{0.2pt}{20pt}   &\rule[-6pt]{0.2pt}{20pt}    \\[0.2cm]
$[J_{a},P_{b}]$&anisotropy $\sim\frac{\Lambda}{\kappa}$ &anisotropy $\sim\frac{\Lambda}{\kappa}$&isotropy\\[0.2cm]
$[J_{a},K_{b}]$& \rule[-6pt]{0.2pt}{20pt}&\rule[-6pt]{0.2pt}{20pt}&\rule[-6pt]{0.2pt}{20pt}\\[0.2cm]
$[J_{a},P_{0}]$&\multicolumn{3}{c}{\rule[2pt]{3.4cm}{0.2pt} $0$ \rule[2pt]{3.4cm}{0.2pt}}\\[0.2cm]
\hline
\\[-0.2cm] 
$[K_{a},K_{b}]$&$O( \frac{\sqrt{\Lambda}}{\kappa})$&$-\epsilon_{abc} J_{c}+O( \frac{\sqrt{\Lambda}}{\kappa})$&$0$ \\[0.2cm]
$[K_{a},P_{b}]$&$O( \frac{\sqrt{\Lambda}}{\kappa}, \frac{1}{\kappa})$&$\delta_{ab} P_{0}+O(\frac{\Lambda}{\kappa}, \frac{1}{\kappa}) $& $\delta_{ab} P_{0}+O(\frac{\Lambda}{\kappa},\frac{1}{\kappa})$\\[0.2cm]
$[K_{a},P_{0}]$&$P_{a}$&$P_{a}$&$0$\\[0.2cm]
\hline
\\[-0.2cm]
$[P_{a},P_{b}]$&$O(\frac{\sqrt{\Lambda}}{\kappa})$&$\Lambda \epsilon_{abc}J_{c}+O(\frac{\sqrt{\Lambda}}{\kappa})$ & $\Lambda \epsilon_{abc}J_{c}$\\[0.2cm]
$[P_{a},P_{0}]$&$\Lambda K_{a}$&$\Lambda K_{a}$&$\Lambda K_{a}$\\[0.2cm]
\hline
\end{tabular}
 \end{center}
}
 \end{table}


\section{Noncommutative spacetimes}\label{sec:ncst}

Besides looking at the properties of the algebra of quantum-deformed relativistic symmetries, it is also instructive to study the properties of the associated noncommutative spacetimes, in which the interplay previously analyzed can be also illustrated. Moreover, since Poincar\'e, (A)dS, Galilei and Carroll classical spacetimes are homogeneous spaces of the corresponding kinematical groups, their noncommutative analogues 
can be constructed as quantum homogeneous spaces of the corresponding quantum groups, although their construction procedure is rather involved from the computational viewpoint (see, for instance,~\cite{DijkhuizenKoornwinder1994,Ciccoli1996qplanes}). Nevertheless, the noncommutative algebra defining a given quantum homogeneous space is just the quantization of the Poisson homogeneous space that is associated to the $r$-matrix defining the first-order of the quantum kinematical algebra. This Poisson homogeneous space is just the classical homogeneous space endowed with a Poisson algebra structure which  can be explicitly obtained as a canonical projection of the Sklyanin Poisson bracket that is derived from the $r$-matrix, provided that the so called coisotropy condition holds (see~\cite{coisotropy}). In the following we will present the explicit expressions for the Poisson-noncommutative spacetimes corresponding to the quantum deformations presented in the previous sections. All technical aspects of this construction as well as appropriate references can be found in~\cite{BGH2019spacetime,BGGH2020kappanewtoncarroll}.

We mentioned when introducing the classical homogeneous spacetimes that their definition requires us to identify the spacetime coordinates as the group parameters of the spacetime translations $P_\alpha$. As we have seen in the previous section (see also Table \ref{table3}), the algebra of translation generators is especially sensitive to the presence of curvature (both with and without quantum deformation). For this reason, we  expect that the same happens to spacetime noncommutativity, and indeed this is the case as shown below.

\subsection{The $\kappa$-(A)dS spacetime}

By  computing the Sklyanin bracket for the $\kappa$-\adsw  $r$-matrix \eqref{r31}  we get the semiclassical version of the $\kappa$-\adsw spacetime in terms of the  Poisson brackets
\begin{align}
\begin{split}
&\{x^0,x^1\} =-\frac{1}{\kappa}\, \frac{\tanh (\eta x^1)}{\eta \cosh^2(\eta x^2) \cosh^2(\eta x^3)} ,\\
&\{x^0,x^2\} =-\frac{1}{\kappa}\,\frac{\tanh (\eta x^2)}{\eta \cosh^2(\eta x^3)} ,\\
&\{x^0,x^3\} =-\frac{1}{\kappa}\,\frac{\tanh (\eta x^3)}{\eta},\\
&\{x^1,x^2\} =-\frac{1}{\kappa}\,\frac{\cosh (\eta x^1) \tanh ^2(\eta x^3)}{\eta} ,\\
&\{x^1,x^3\} =\frac{1}{\kappa}\,\frac{\cosh (\eta x^1) \tanh (\eta x^2) \tanh (\eta x^3)}{\eta} ,\\
&\{x^2,x^3\} =-\frac{1}{\kappa}\,\frac{\sinh (\eta x^1) \tanh (\eta x^3)}{\eta} \, ,
\end{split}
\label{eq:kAdSST}
\end{align} 
where we defined $\eta^2=-\Lambda$ so that the zero-curvature limit, giving the $\kappa$-Minkwoski Poisson homogeneous space (whose quantization is the $\kappa$-Minkowski noncommutative spacetime~\eqref{kminkowski})  is given by the $\eta\to 0$ limit of~\eqref{eq:kAdSST}, namely:
\begin{align}
\begin{split}
\label{eq:PoissonkappaMinkowski}
&\{x^0,x^a\}=- \frac{1}{\kappa} \, x^a, \qquad  \{x^a,x^b\}= 0\, ,
\end{split}
\end{align} 
and in this flat case space translations do commute among themselves.  Indeed, if we take the first-order expansion in terms of $\eta$ we get 
\begin{align}
\begin{split}
&\{x^0,x^1\} =- \frac{1}{\kappa}\, (x^1 + o[\eta^2]), \\
&\{x^0,x^2\} =- \frac{1}{\kappa}\, (x^2 + o[\eta^2]), \\
&\{x^0,x^3\} =- \frac{1}{\kappa}\, (x^3 + o[\eta^2]) ,\\
&\{x^1,x^2\} =-\frac{1}{\kappa}\,(\eta\,(x^3)^2 + o[\eta^2]), \\
&\{x^1,x^3\} =\frac{1}{\kappa}\,(\eta\,x^2 x^3 + o[\eta^2]),\\
&\{x^2,x^3\} =-\frac{1}{\kappa}\,(\eta\,x^1 x^3 + o[\eta^2]).
\end{split}
\end{align} 
Notice that curvature has a more prominent role in the space-space brackets, where it contributes at the order $\frac{\sqrt{\Lambda}}{\kappa}$, while for the time-space brackets curvature  only contributes starting from the $\frac{{\Lambda}}{\kappa}$ order. This behavior is similar (but not completely equal) to the properties of the algebra of translation generators, schematically described in Table \ref{table3}. In fact, the quantum-curvature effects in the commutators between space-space generators are governed by $O(\frac{\sqrt{\Lambda}}{\kappa})$ (similarly to what happens to the brackets between spatial coordinates), while for time-space commutators one has no contributions at all from quantum effects (for the time-space coordinates there is a contribution, even though it is milder than in the space-space case). The quantization of the $\kappa$(A)dS Poisson homogeneous spacetime was fully given in~\cite{BGH2019spacetime} by choosing a precise ordering of the generators, but the interplay between $\Lambda$ and $\kappa$ here presented is not modified after quantization. We recall that other noncommutative (A)dS spacetimes arising from different noncommutative geometry approaches can be found in~\cite{Stein,BM,VH,Manolakos:2019fle}.

\subsection{$\kappa$-Galilean and $\kappa$-Carrollian spacetimes}

The Galilean and Carrollian limits of the $\kappa$-(A)dS spacetime \eqref{eq:kAdSST} are obtained by appropriately rescaling spacetime coordinates to so keep the  products $x^0P_0$ and $x^aP_a$  invariant under  contraction (see~\cite{BHOS1995jmp} for the theory of contractions of Poisson-Lie groups and~\cite{BGGH2020kappanewtoncarroll}, where these two limits have been performed onto the Snyder noncommutative spacetime~\cite{Snyder1947}).

Specifically,  the Galilean limit is obtained by rescaling
\be
x^{a}\rightarrow c \,x^{a}\,,
\ee
and then taking the $c\rightarrow\infty$ limit of \eqref{eq:kAdSST}. This produces a spacetime algebra which has the same commutation rules as $\kappa$-Minkowski for the space-time sector, and shows the residual anisotropy discussed above in section \ref{sub:kAdScontractions} in the space sector:
\be
 \{x^a,x^0\} =\frac{1}{\kappa}\, x^a ,\quad\  \{x^1,x^2\} =-\frac{\eta}{\kappa}\,   (x^3)^2    ,\quad\  \{x^1,x^3\} =\frac{\eta}{\kappa}\,  x^2 x^3
 ,\quad\  \{x^2,x^3\} =-\frac{\eta}{\kappa}\,  x^1 x^3 .
\ee
Symplectic leaves for the space sector are just 3-spheres
\be
S=(x^1)^2 + (x^2)^2 + (x^3)^2 \, ,
\label{spheres}
\ee
which reflects the role of the deformed SO(3) sector~\eqref{deformedso3} in both $\kappa$-(A)dS and curved (Newton-Hooke) $\kappa$-Galilean algebras and spaces.

The Carrollian limit is obtained as the limit $c\rightarrow 0$ of \eqref{eq:kAdSST}, after the following rescaling is performed (notice that, as done for the algebra of symmetries, the quantum deformation parameters has to be also rescaled)
\be
x^{0}\rightarrow  x^{0}/c \,,\qquad \kappa\rightarrow c\,\kappa\,.
\ee
In this case, the space-time part of the algebra is not affected by the contraction, and remains equal to the one of $\kappa$-(A)dS. The most important effect of the contraction is the restoration of isotropy at the level of spatial coordinates, consistently with what found  in section \ref{sub:kAdScontractions} at the level of the algebra of symmetries:
\begin{align}
\begin{split}
\label{di}
&\{x^1,x^0\} =\frac{1}{\kappa}\, \frac{\tanh (\eta x^1)}{\eta \cosh^2(\eta x^2) \cosh^2(\eta x^3)} ,\\
&\{x^2,x^0\} =\frac{1}{\kappa}\,\frac{\tanh (\eta x^2)}{\eta \cosh^2(\eta x^3)} ,\\
&\{x^3,x^0\} =\frac{1}{\kappa}\,\frac{\tanh (\eta x^3)}{\eta},\\
&\{x^a,x^b\} =0.
\end{split}
\end{align} 

When the flat $\Lambda\rightarrow 0$ limit is taken, in both cases one recovers the same $\kappa$-Minkowski Poisson algebra~\eqref{eq:PoissonkappaMinkowski}. In particular, as seen for the associated algebra of symmetries in section \ref{sub:kAdScontractions}, isotropy is restored also in the Galilean case. As it can be described in~\cite{BGGH2020kappanewtoncarroll}, the quantization of all these Galilean and Carrollian Poisson homogeneous spacetimes can be fully performed by mimicking the quantization procedure used in the $\kappa$-(A)dS case. In particular, in the curved Galilean case the ``quantum spheres"
 \be
\hat S_{\eta/\kappa}=(\hat x^1)^2 + (\hat x^2)^2 + (\hat x^3)^2  + \frac{\eta}{\kappa}\, \hat x^1 \hat x^2,\qquad [\hat S_{\eta/\kappa} , \hat x^a]=0,
\label{df}
\ee
are obtained as the quantization of the symplectic leaves~\eqref{spheres}, where the  term depending on $\eta/\kappa$ arises from the noncommutativity of the quantum space coordinates $\hat x^a$.


\section{Concluding remarks}\label{sec:conclusions}

There exist two more frameworks in which the results here presented for each of the quantum kinematical algebras and their associated noncommutative spacetimes can be rephrased. 

Firstly, all the models here presented could be analyzed in terms of the associated curved momentum spaces.
These are pseudo-Riemannian manifolds that can be obtained  as orbits of suitable actions of the dual Poisson-Lie group associated to the $\kappa $-deformation. In the case of $\kappa$-Poincar\'e, as was first shown in~\cite{KowalskiGlikman:2002ft}, the geometry one finds is that of one half of de Sitter space. This analysis can be generalized to more general $\kappa$-deformations of the $ISO(p,q)$ group and its Carrollian contractions, in which the ``deformed'' direction is not necessarily the ``time'' one (the zeroth coordinate). The result is a collection of 4-dimensional momentum spaces which always have the geometry of a homogeneous space (dS, AdS or Minkowski), and in some cases cover only half of said geometries, in other case covers a whole sheet (as in the Euclidean case $ISO_\kappa(4)$~\cite{Lizzi:2020tci}).

In the case of $\kappa$-(A)dS, the Lie bialgebra~(\ref{cocoads}) dualizes to a Lie algebra which admits a 7-dimensional solvable Lie subalgebra that includes the duals of the translation and boost generators. Therefore the smallest generalization of momentum space is 7-dimensional generalization of momentum space, which includes three additional coordinates associated to ``hyperbolic angular momentum"~\cite{Ballesteros:2017pdw}. The geometry of these momentum spaces is half of the $(6+1)$-
dimensional de Sitter space in the case of $\kappa$-dS, and half of a space with $SO(4, 4)$  invariance for  $\kappa$-(A)dS. The Galilean and Carrollian limits of these momentum spaces have not been studied yet, and are worth further investigation.

Secondly, an alternative viewpoint is provided by the construction of the corresponding noncommutative spaces of worldlines associated to all the $\kappa$-deformations here presented. In particular, for the (A)dS and Poincar\'e cases, the spaces of time-like worldlines are obtained as homogeneous spaces of cosets of the corresponding Lie group with respect to the 4D isotropy subgroup of the worldline of a particle located at the origin of the spacetime and having zero velocity, which is generated by the subalgebra of symmetries given by ${\frak h}=\{J_1,J_2,J_3,P_0 \}$ (see~\cite{BGH2019worldlinesplb} and references therein).

In the Poincar\'e case the classical 6D space of time-like worldlines ${\cal W}$ has been explicitly constructed, and has been endowed with a Poisson homogeneous structure associated to the $\kappa$-Poincar\'e $r$-matrix \eqref{kPr}. As it was shown in~\cite{BGH2019worldlinesplb}, this structure provides a Poisson algebra on the space of worldlines coordinates, that can be quantized giving rise to the quantum space of worldlines associated to the $\kappa$-Poincar\'e symmetry. This noncommutative space of time-like worldlines provides an alternative (and physically sound) framework for the description of the spacetime fuzziness encoded in quantum deformations~\cite{FuzzyWorldlines}. The construction of the noncommutative spaces of worldlines associated to the $\kappa$-(A)dS, $\kappa$-Galilean and $\kappa$-Carrollian algebras can be attempted by adopting a similar approach, thus providing a complementary perspective for the analysis of the interplay between the quantum deformation parameter $\kappa$, the curvature parameter $\Lambda$ and the speed of light $c$.


\section*{Acknowledgements}

This work has been partially supported by Ministerio de Ciencia e Innovaci\'on (Spain) under grants MTM2016-79639-P (AEI/FEDER, UE) and PID2019 - 106802GB-I00 / AEI / 10.13039 / 501100011033, by Junta de Castilla y Le\'on (Spain) under grants BU229P18 and BU091G19, as well as by the Action CA18108 QG-MM from the European Cooperation in Science and Technology (COST). 




\end{document}